\documentclass[prb,twocolumn,amsmath,superscriptaddress,amssymb]{revtex4-1}
\usepackage[section]{placeins}
\usepackage{setspace}
\usepackage{graphicx}
\usepackage{amsmath,amssymb}
\usepackage{amsbsy}
\usepackage{color}
\usepackage{multirow}
\usepackage[latin1]{inputenc}
\usepackage[T1]{fontenc}
\usepackage[english]{babel}
\usepackage{amsmath,amssymb,amsthm}
\usepackage{hyperref}
\usepackage{booktabs}
\usepackage{bbm}
\usepackage{url}
\usepackage{nicefrac}
\usepackage{multirow} 
\usepackage{graphicx}
\usepackage{array}
\usepackage{times}

\newcommand{\qo}[1]{``#1''}

\usepackage{array}
\renewcommand{\epsilon}{\varepsilon}
\renewcommand{\phi}{\varphi}

\newcolumntype{L}[1]{>{\raggedright\let\newline\\\arraybackslash\hspace{0pt}}m{#1}}
\newcolumntype{C}[1]{>{\centering\let\newline\\\arraybackslash\hspace{0pt}}m{#1}}
\newcolumntype{R}[1]{>{\raggedleft\let\newline\\\arraybackslash\hspace{0pt}}m{#1}}
\usepackage{sistyle}
\usepackage{soul}
\definecolor{lightblue}{RGB}{185,210,248}
\sethlcolor{lightblue}

\begin{document}
\title{High-dimensional quantum cloning and applications to quantum hacking}

\author{Fr\'ed\'eric Bouchard}
\affiliation{The Max Planck Centre for Extreme and Quantum Photonics, Department of Physics, University of Ottawa, 25 Templeton St., Ottawa, Ontario, K1N 6N5 Canada}
\author{Robert Fickler}
\affiliation{The Max Planck Centre for Extreme and Quantum Photonics, Department of Physics, University of Ottawa, 25 Templeton St., Ottawa, Ontario, K1N 6N5 Canada}
\author{Robert W. Boyd}
\affiliation{The Max Planck Centre for Extreme and Quantum Photonics, Department of Physics, University of Ottawa, 25 Templeton St., Ottawa, Ontario, K1N 6N5 Canada}
\affiliation{Institute of Optics, University of Rochester, Rochester, New York, 14627, USA}
\author{Ebrahim Karimi}
\email{ekarimi@uottawa.ca}
\affiliation{The Max Planck Centre for Extreme and Quantum Photonics, Department of Physics, University of Ottawa, 25 Templeton St., Ottawa, Ontario, K1N 6N5 Canada}
\affiliation{Department of Physics, Institute for Advanced Studies in Basic Sciences, 45137-66731 Zanjan, Iran}

\begin{abstract}
Attempts at cloning a quantum system result in the introduction of imperfections in the state of the copies. This is a consequence of the no-cloning theorem, which is a fundamental law of quantum physics and the backbone of security for quantum communications. Although such perfect copies are prohibited, a quantum state may be copied with maximal accuracy via various optimal cloning schemes. Optimal quantum cloning, which lies at the border of the physical limit imposed by the no-signalling theorem and the Heisenberg uncertainty principle, has been experimentally realized for low dimensional photonic states. However, an increase in the dimensionality of quantum systems is greatly beneficial to quantum computation and communication protocols. Nonetheless, no experimental demonstration of optimal cloning machines has hitherto been shown for high-dimensional quantum systems. Here, we perform optimal cloning of high-dimensional photonic states by means of the symmetrization method. We show the universality of our technique by conducting cloning of numerous arbitrary input states, and fully characterize our cloning machine by performing quantum state tomography on \emph{cloned} photons. In addition, a cloning attack on a  Bennett and Brassard (BB84) quantum key distribution protocol is experimentally demonstrated in order to reveal the robustness of high-dimensional states in quantum cryptography.
\end{abstract}
\maketitle 

High-dimensional information is a promising field of quantum information science that has been strongly matured over the last years. It is known that by using not only \textit{qubits} but \textit{qudits}, i.e. $d$-dimensional quantum states, it is possible to encode more information on a single carrier, increase noise-resistance in quantum cryptography protocols~\cite{cerf:02}, and investigate fundamentals of nature~\cite{krenn:14}. Photonic systems have shown to be promising candidates in quantum computation and cryptography for many proof-of-principle demonstrations as well as for \qo{flying} quantum carrier to distribute high-dimensionally encoded states. Orbital angular momentum (OAM) of light, which provides an unbounded vector space, has long been recognized as a potential high-dimensional degree of freedom for conducting experiments on foundations of quantum mechanics~\cite{dada:11,harris:16}, quantum computation~\cite{cardano:15} and cryptography~\cite{mirhosseini:15}. The main characteristic of photons carrying OAM is their twisted wavefront characterized by $\exp(i \ell \phi)$ phase term, where $\ell$ is an integer and $\phi$ is the azimuthal coordinate~\cite{allen:92}. In the context of quantum information, OAM states of photons have the particular advantage of representing quantum states belonging to an infinitely large, but discrete, Hilbert space~\cite{molina:07}. Finite sub-spaces of dimensions $d$ can be considered as laboratory realizations of photonic qudits. In this study, we adopt the OAM degree of freedom of single photons to achieve high-dimensional quantum cloning and performing quantum hacking on a high-dimensional quantum communication channel.
Although perfect cloning of unknown quantum states is forbidden~\cite{wootters:82}, it is interesting to ask `how similar to the initial quantum state the best possible quantum clone can be?' The answer is given in terms of the cloning fidelity ${\cal F}$, which is defined as the overlap between the initial state to be cloned and that of the cloned copies. This figure of merit is a significant measure of the accuracy of a cloned copy obtained from a specific cloner. Schemes that achieve the best possible fidelity are called optimal quantum cloning, and play an important role in quantum information~\cite{gisin:97}. For instance, an optimal state estimation yields a bounded fidelity of ${\cal F}_\mathrm{est}=2/(1+d)$, where $d$ is the dimension of the quantum state~\cite{Bruss:99}. Optimal quantum cloning turns out to be a more efficient way of broadcasting the quantum state of a single system as it yields a fidelity that is always higher than that of optimal state estimation, which has been experimentally realized for low dimensional photonic states~\cite{lamas:02,irvine:04,nagali:09,nagali:10}. Moreover, this enhancement in fidelity grows larger with higher dimensional quantum states, further motivating experimental investigations of high-dimensional quantum cloning. Hence, high-dimensional optimal quantum cloning machines are of great importance whenever quantum information is to be transmitted among multiple individuals without knowledge of the input quantum state. Here, we concentrate on the $1 \rightarrow 2$ universal optimal quantum cloning machine, for which the optimal fidelity of the two cloned copies is given by ${\cal F}_\mathrm{clo}=1/2+1/(1+d)$, where $d$ is the dimension of the Hilbert space of the states that are to be cloned~\cite{Navez:03}.

%
\begin{figure}[!htbp]
	\begin{center}
	\includegraphics[width=0.9\columnwidth]{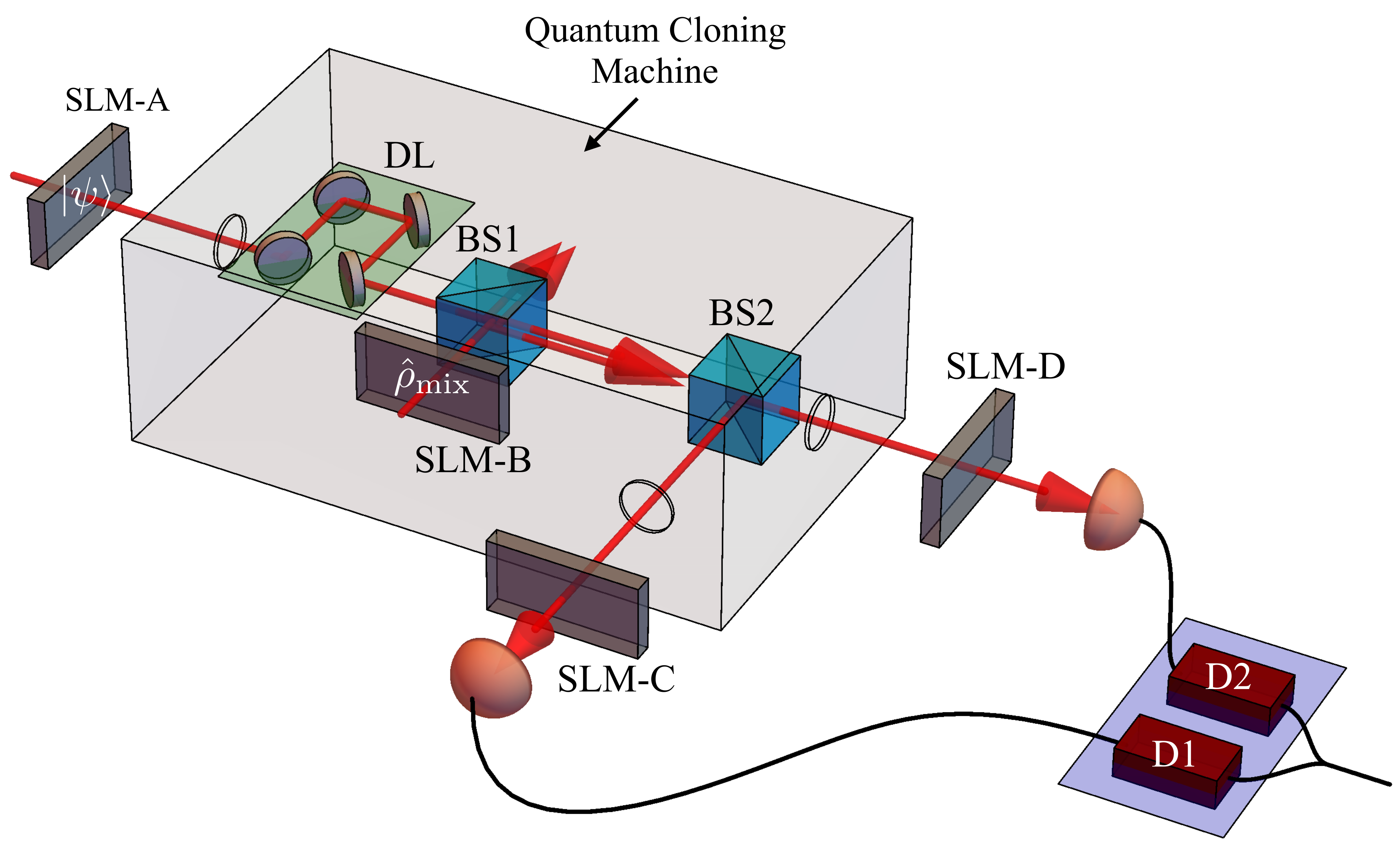}
	\caption[]{Simplified sketch of the experimental setup. The input quantum state $|\psi \rangle$ is imprinted on a single photon using a spatial light modulator (SLM-A). The single photon is subsequently sent to the cloning machine for optimal cloning. The cloning machine consist of a delay line (DL), to adjust the arrival time of the input photon, a second photon that is in a completely mixed state when exiting SLM-B and a first beam splitter (BS1). The two photons are made to arrive at the beam splitter simultaneously using the DL. The two photons exiting one of the output port of the first beam splitter together are further separated at a second beam splitter (BS2), and are sent out of the cloning machine. The cloned photons are then detected and characterized using detectors (D1 and D2) and spatial light modulators (SLM-C and SLM-D), respectively. }
	\label{fig1}
	\end{center}
\end{figure}

We use the symmetrization method to realize a universal optimal quantum cloning machine for high-dimensional OAM states~\cite{ricci:04,sciarrino:04}. In this method, the quantum state that is to be cloned, namely $| \psi \rangle$, is sent in one of the input port of a non-polarizing beam splitter. In the other input port, a completely mixed state of the appropriate dimension, given by $\hat{\rho}_\mathrm{mix}=I_d/d$, is sent, where $I_d$ is the $d$-dimensional identity matrix. Therefore, when both input photons are interfering at the beam splitter, two \qo{cloned} photons will jointly exit one of the output ports. We note that this cloning scheme does not require knowledge of the input state and applies to any arbitrary state. Indeed, this property is a result of the \qo{universality} of the cloning machine and shows the versatility of our scheme. Each output cloned photon's state is represented by a reduced density matrix obtained by tracing over the other photon. Since both cloned photons are characterized by an identical cloned state, the cloner is then said to be \qo{symmetric}. Hence, the symmetrization method is considered to be a symmetric optimal universal quantum cloning machine (UQCM). In our experiment we implement a high-dimensional version of this UQCM with OAM states of single photons (See Fig.\ref{fig1}). We generate and measure the OAM states by manipulating the phase front of the photons using a liquid-crystal phase-only spatial light modulator~\cite{SM}.

%
\begin{figure}[!htbp]
	\begin{center}
	\includegraphics[width=1\columnwidth]{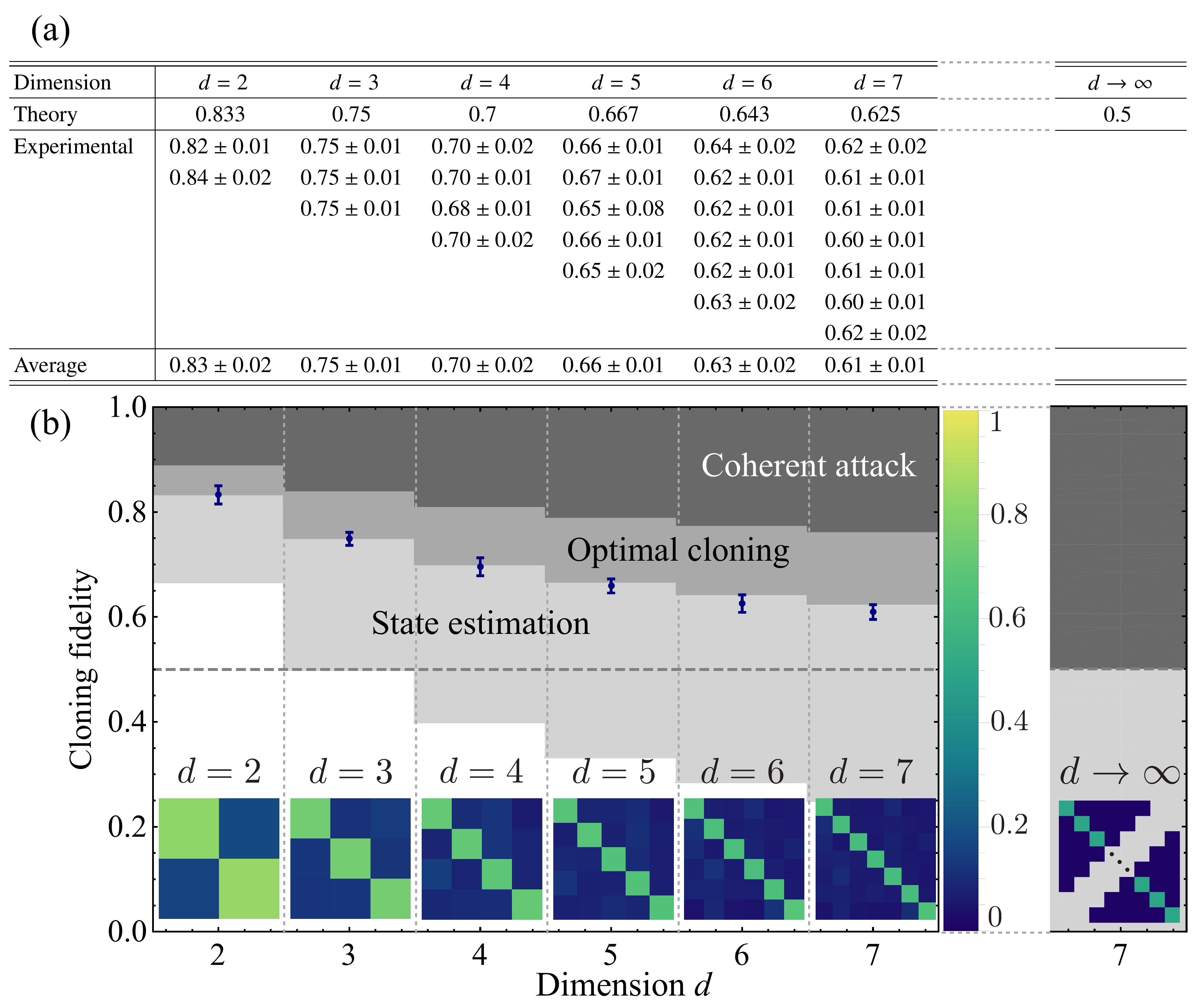}
	\caption[]{Optimal cloning fidelity for various dimensions. (a) Experimental values of the cloning fidelities are shown for each $d$ number of elements of the logical basis, along with theoretical values, for various dimensions $d$. (b) The average cloning fidelities (blue dots) are plotted for various dimensions, along with probability matrices ${\cal P} \left( | i \rangle , | \psi \rangle \right)$ of detecting a cloned photon in any output state $|i\rangle$ of the OAM logical basis, given an input state $|\psi\rangle$ of the same basis. The diagonal elements of the probability matrices corresponds to the cloning fidelity of each elements of the basis. The light and dark grey shaded area corresponds to fidelities not accessible by state estimation and $1 \rightarrow 2$ optimal symmetric UQCM, respectively. In quantum cryptography, a more effective class of quantum hacking, namely coherent attacks~\cite{cerf:02}, yields larger fidelities illustrated by the dim grey shaded area.}
	\label{fig2}
	\end{center}
\end{figure}

In order to characterize the quality of our UQCM, we use two different approaches to evaluate the yielding cloning fidelities: measuring the probability of successful cloning as well as full state tomography of the cloned photons. In this first series of measurements, we evaluate the cloning fidelity, ${\cal F}_\psi$, of a given arbitrary input state, $|\psi \rangle$, from the probability of finding both output cloned photons in the state $| \psi \rangle$, i.e. ${\cal P} \left( |\psi\rangle , | \psi \rangle \right)$. This probability can be obtained experimentally by means of coincidence measurements: ${\cal F}_{\psi} ={\cal P} \left( |\psi\rangle , | \psi \rangle \right)=\left( N\left( |\psi \rangle,| \psi \rangle \right)+\sum_{i \neq \psi} N\left( |\psi \rangle,| i \rangle \right) \right)/N_{\mathrm{tot}}$, where $N\left( |i\rangle, |j \rangle \right)$ represents the number of coincidence measurements between the states $| i \rangle$ and $|j\rangle$, $N_{\mathrm{tot}}$ is the total number of coincidence measurements, i.e. $N_\mathrm{tot}=N\left( |\psi \rangle,| \psi \rangle \right)+2\sum_{i \neq \psi} N\left( |\psi \rangle,| i \rangle \right)$, and $|i \rangle$ and $|j \rangle$ represent elements of the basis containing $|\psi \rangle$. The factor of 2 that appears in the definition of $N_\mathrm{tot}$ is a result of the symmetric nature of our cloning machine, where $N\left( | i \rangle, | j \rangle \right)=N\left( | j \rangle, | i \rangle \right)$. Further, one can obtain from normalization, ${\cal P} \left( | i \rangle , | \psi \rangle \right)=N\left( | i\rangle, | \psi \rangle \right)/N_\mathrm{tot}$, for $i \neq \psi$.

\begin{figure*}[!htbp]
	\begin{center}
	\includegraphics[width=2\columnwidth]{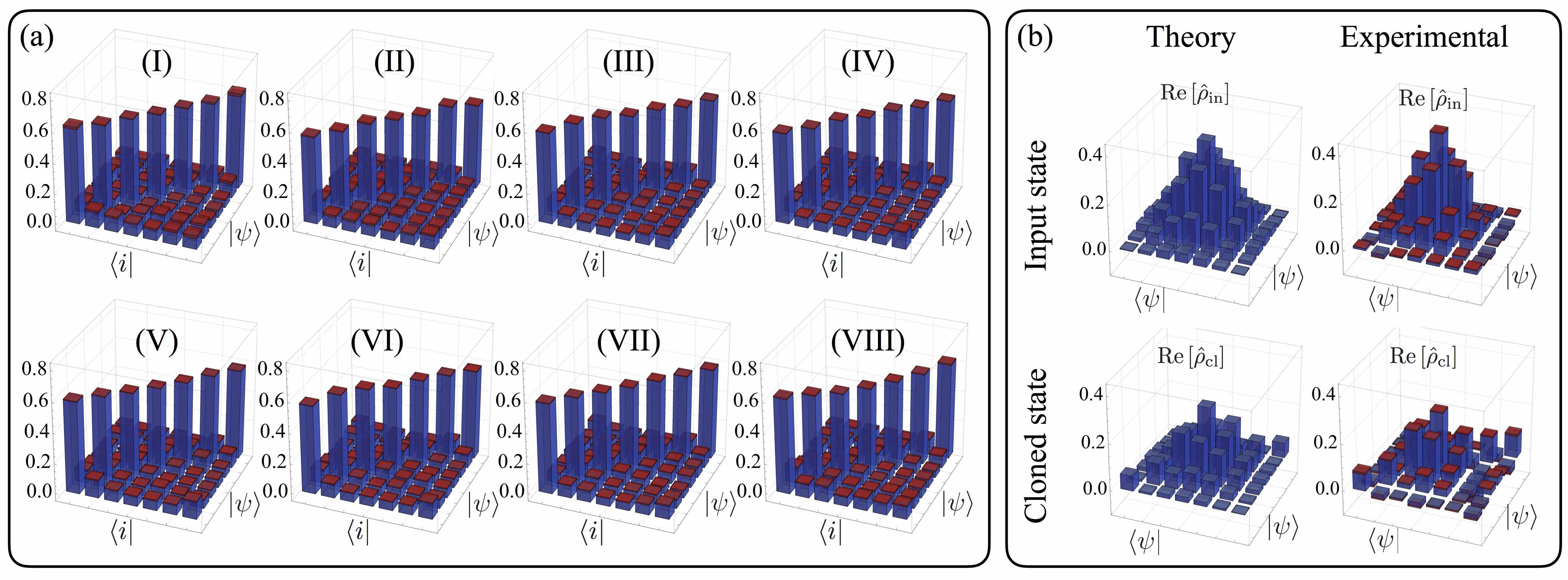}
	\caption[]{Cloning fidelities for various MUBs and cloning of gaussian states in dimension $d=7$. (a) Probability ${\cal P} \left( | i \rangle , | \psi \rangle \right)$ of detection of an output cloned state $|i\rangle$ given an input state $|\psi\rangle$, where $|i\rangle$ and $|\psi \rangle$ belong to a specific MUB. This set of measurement is repeated for all $d+1$ MUBs (I)-(VIII), in dimension 7. The on-diagonal elements represent the cloning fidelities for each element of a given basis. (b) Theoretical and experimental high-dimensional cloning of a gaussian state. The cloned fidelity is obtained by calculating the overlap of the reduced density matrix of the cloned state with the input state. The experimental reduced density matrix of the cloned state is obtained by full quantum state tomography. The experimentally reconstructed density matrices of the Gaussian state before and after cloning are shown along with their theoretical counterpart.}
	\label{fig3}
	\end{center}
\end{figure*}

Compared to a full tomographic reconstruction, this method requires fewer measurements and thus enables us to characterize the cloning fidelity of our cloner under a wider range of circumstances. For instance, the effect of dimensionality on a UQCM is a crucial point for any optimal cloning schemes. As mention previously, increasing the dimensionality of the input quantum states results in a decrease of the cloning fidelity. Interestingly, this decrease in cloning fidelity can serve as an intuitive explanation of the superiority of high-dimensional quantum cryptography. In our experiment we measure the cloning fidelity of our cloning machine for different input states belonging to the computational OAM basis of various dimensions $d \in \{2,3,4,5,6,7 \}$. We find near-perfect agreement of the experimentally evaluated cloning fidelities to the theoretical predictions of the high-dimensional $1\rightarrow2$ symmetric optimal UQCM (see Fig~\ref{fig2}). In addition, we experimentally verify the universality of our cloning machine by performing quantum cloning of every states of all $d+1$ OAM mutually unbiased bases (MUBs)~\cite{durt:10,klappenecker:04} (See Fig~\ref{fig3}-(a)). Once more, we find near optimal cloning fidelities for all MUBs, thus demonstrating the viability and universality of our optimal quantum cloner (See supplementary information). Note that MUBs and their elements are playing an important role in quantum communication and information, e.g. as basis states utilized in quantum cryptographic protocols~\cite{bennett:84} or quantum state tomography~\cite{adamson:10}.

As a second series of measurements for a complete characterization of our UQCM, we fully reconstruct the high-dimensional cloned quantum states by means of quantum state tomography. Moreover, our UQCM should be able to clone the state regardless of the input state and its complex structure in the high-dimensional state space. An exemplary and visually interesting high-dimensional state is the so-called \textit{Gaussian} state given by the following superposition, $|\psi_\mathrm{Gauss} \rangle= { \cal N} \sum_{\ell=-3}^{\ell=3} \exp \left[ - \left( \ell/2 \right)^2 \right] | \ell \rangle$, where ${\cal N}$ is a normalization constant. We experimentally generate the Gaussian state of dimension $d=7$ and perform full quantum state tomography on one of the output cloned photons. The theoretically expected and experimentally achieved results are shown in Fig.~\ref{fig3}-(b). The cloned Gaussian state has a fidelity $0.80 \pm 0.03$ with respect to the theoretically expected cloned density matrix. Thus, for an arbitrary complex input state $| \psi_\mathrm{Gauss} \rangle$ the experimental cloning fidelity of our UQCM obtained from complete quantum state tomography can be estimated to be around $0.40 \pm 0.01$. In comparison to the fidelity for $d=7$ of 0.625, which we evaluated from success probabilities, the lower value can be explained by preparation and detection errors, since full quantum state tomography requires a much larger number of measurements. However, both methods show that our implementation of a symmetric UQCM can be used to clone any arbitrarily complex quantum state up to dimension seven, without a significant loss or deterioration of the optimal state fidelities. Hence, cloning of high-dimensional quantum states encoded in the OAM degree of freedom might become a building block of future high-dimensional quantum information science.

%
\begin{figure}[!htbp]
	\begin{center}
	\includegraphics[width=0.9\columnwidth]{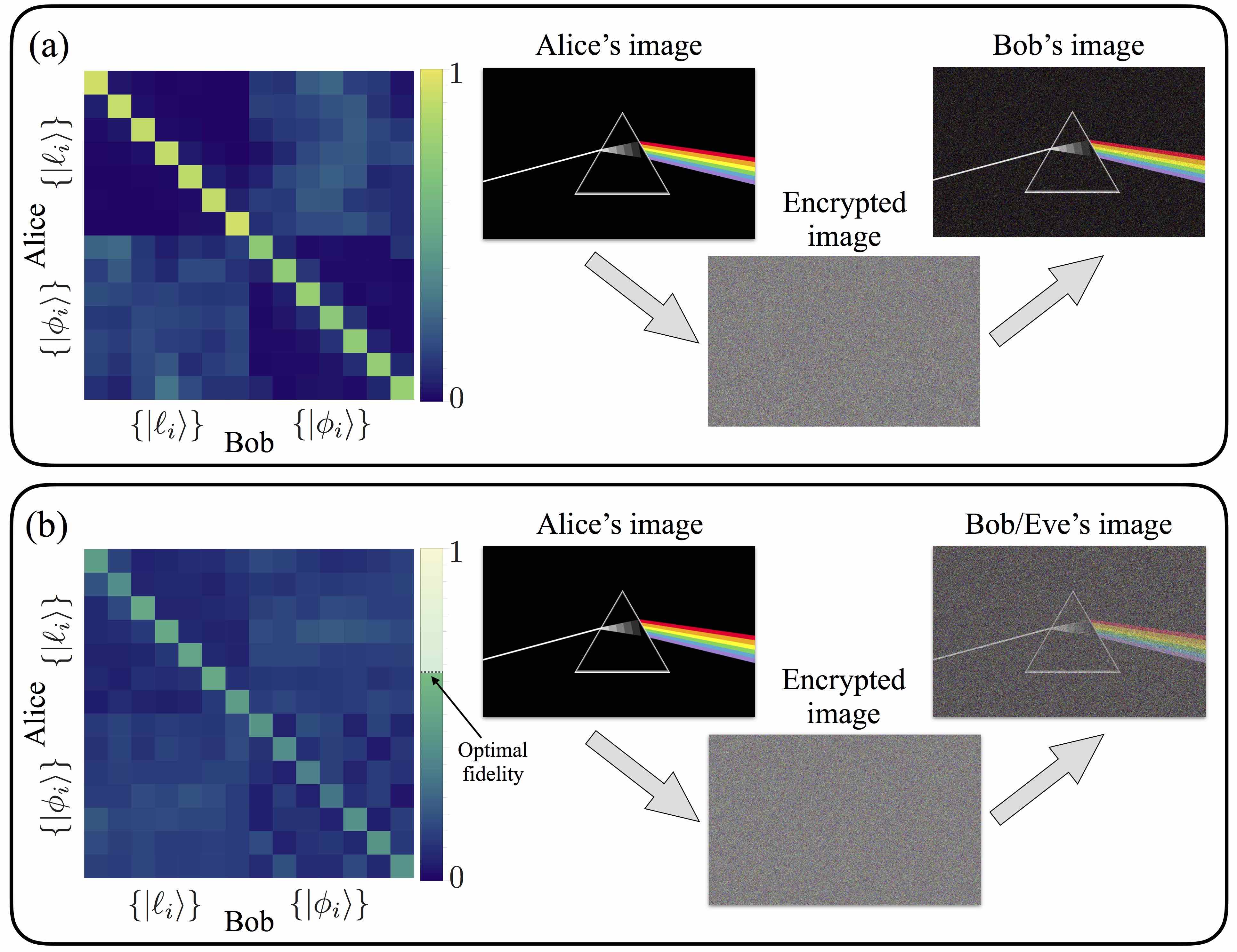}
	\caption[]{High-dimensional quantum key distribution with and without quantum hacking. (a) Experimental probability matrices obtained from projective measurements are shown on the left side. The bases selected by Alice and Bob are indicated on the vertical and horizontal axes, respectively. On the right, we show Alice's initial message and Bob's decrypted message. (b) Experimental probability matrices with the presence of an eavesdropper having access to a symmetric optimal UQCM. Similarly, Alice's initial message is shown along with the decrypted message obtained both by Bob and Eve. One may note that for the BB84 protocol, the symmetric UQCM does not lead to the optimal individual attack. Rather, our UQCM results in the optimal individual attack for the QKD protocol exploiting all $d+1$ available MUBs. Indeed, in the simpler case of the BB84 protocol, the optimal attack consists of the asymmetric Fourier-covariant cloner~\cite{cerf:02,fan:03}, which cannot be straightforwardly implemented in our experimental setup.}
	\label{fig4}
	\end{center}
\end{figure}
%

As a final test of the abilities to clone high-dimensional quantum states, we implement a cloning attack into a high-dimensional quantum key distribution (QKD) scheme. In a QKD protocol, a sender (Alice) and receiver (Bob) use quantum states to distribute a random, secret key shared between both parties. The shared key is then use to communicate an encrypted message through a classical channel, using the perfectly secure one-time pad protocol. The security of QKD resides in the fact that the presence of an eavesdropper (Eve) will result in the introduction of errors in the shared key, which can originate for example from the non-perfect but optimal cloning done by the eavesdropper. Note that the dimensionality of the quantum states used to distribute the key affects directly the cloning fidelity and thus the amount of errors introduced by a possible cloning attack. 

We first perform a high-dimensional QKD using the seminal BB84 protocol~\cite{bennett:84}, extended using OAM states of dimension $d=7$, and then experimentally replicate the presence of an eavesdropper, having access to a high-dimensional optimal quantum cloning machine, performing an individual attack. In our experiment, the first MUB is given by the logical OAM basis $\{ |\ell \rangle ; \ell=-3,-2,-1,0,1,2,3 \}$ and the second MUB is given by the Fourier angle basis $\{|\phi_i \rangle ; i=1,2,3,4,5,6,7 \}$~\cite{SM}. Projective measurements are shown with and without the cloning attack in Fig.~\ref{fig4}-(a) and (b), respectively. The lower fidelity due to a cloning attack is readily visible. A visually compelling illustration of the effect of an eavesdropper on Alice's and Bob's shared key can be given by directly using the established raw sifted key, without further performing error correction and privacy amplification, as a one-time pad to share an encrypted message, e.g. an image of their favourite optical phenomenon. We experimentally simulate such a situation by performing the high-dimensional BB84 protocol with and without Eve's attack by using our UQCM. In a real world QKD, experimental errors will always be introduced in the raw key, leading to a slightly deteriorated image after Bob's decryption (see Fig~\ref{fig4}-(a)). However, if Eve performs her cloning attack while Alice and Bob are trying to establish their key, the errors increase significantly which is then directly visible in Bob's decrypted image (see Fig~\ref{fig4}-(b)). The quantum bit error rate (QBER) is given by 0.16 and 0.57, without and with the cloning attack, respectively. In the absence of an eavesdropper, the QBER is well below the error bound for security in dimension 7, i.e. $D_\mathrm{coh}(7)=23.72\%$~\cite{cerf:02}. Thus, error correction and privacy amplification may be performed in order for Alice and Bob to obtained a completely secure and errorless shared key. However, in the presence of the eavesdropper, the QBER is well above the bound in dimension 7, hence revealing immediately the presence of Eve. Furthermore, the mutual information between Alice and Bob may be calculated from $I_{AB}^d = \log_2 (d) + \left(1-e_B^N \right) \log_2 \left(1-e_B^N \right) + e_B^N \log_2 \left( e_B^N / (d-1) \right)$, where $e_B^N$ is Bob's error rate~\cite{zhang:07}. Experimental values of 1.73 and 0.36 bits per photons were obtained for Alice and Bob's mutual information with and without the cloning attack, respectively. Moreover, we performed quantum hacking to a 2-dimensional QKD protocol (BB84). In this case, the quantum bit error rate is given by, 0.19 and 0.007, with and without the cloning attack, which is well above and below the security bound in dimension 2, i.e. $D_\mathrm{coh}(2)=11.00\%$, respectively. Hence, it is clear that high-dimensional quantum cryptography leads to higher signal disturbance in the presence of an optimal cloning attack resulting in a larger tolerance to noise in the quantum channel.

In conclusion, we showed the feasibility of high-dimensional optimal quantum cloning of orbital angular momentum states of single photons. This scheme was further employed to perform a cloning attack to a secure quantum channel, revealing the robustness of high-dimensional quantum cryptography upon quantum hacking. Moreover, studying the effect of dimensionality and universality on optimal quantum cloning reveals its advantage over optimal state estimation in quantum information schemes where unknown quantum states must be distributed.


\bibliographystyle{Science}

%
%
%
%
%
%
%
%
%
%
%
\vspace{1cm}
\noindent\textbf{Supplementary Materials} Materials and Methods, Supplementary Text, and Figs.~\ref{figsm1} and \ref{figsm2}.
\vspace{1 EM}

\noindent\textbf{Acknowledgments}
\noindent This work is supported by the Canada Research Chairs (CRC) Program and Canada Foundation for Innovation (CFI). F.B. acknowledges the support of the Vanier Canada Graduate Scholarships Program and the Natural Sciences and Engineering Research Council of Canada (NSERC) Canada Graduate Scholarships program. R.F. acknowledges the support of the Banting postdoctoral fellowship of the NSERC.
\vspace{1 EM}

\clearpage
\setcounter{figure}{0} \renewcommand{\thefigure}{S\arabic{figure}}
\setcounter{table}{0} \renewcommand{\thetable}{S\arabic{table}}
\setcounter{section}{0} \renewcommand{\thesection}{S\arabic{section}}
\setcounter{equation}{0} \renewcommand{\theequation}{S\arabic{equation}}
\onecolumngrid
\section*{{\Large Supplementary Materials for}\\ High-dimensional quantum cloning and applications to quantum hacking}

\section*{\underline{\large{Part 1}}\\ Method:}
\begin{figure}[!htbp]
	\begin{center}
	\includegraphics[width=0.95\columnwidth]{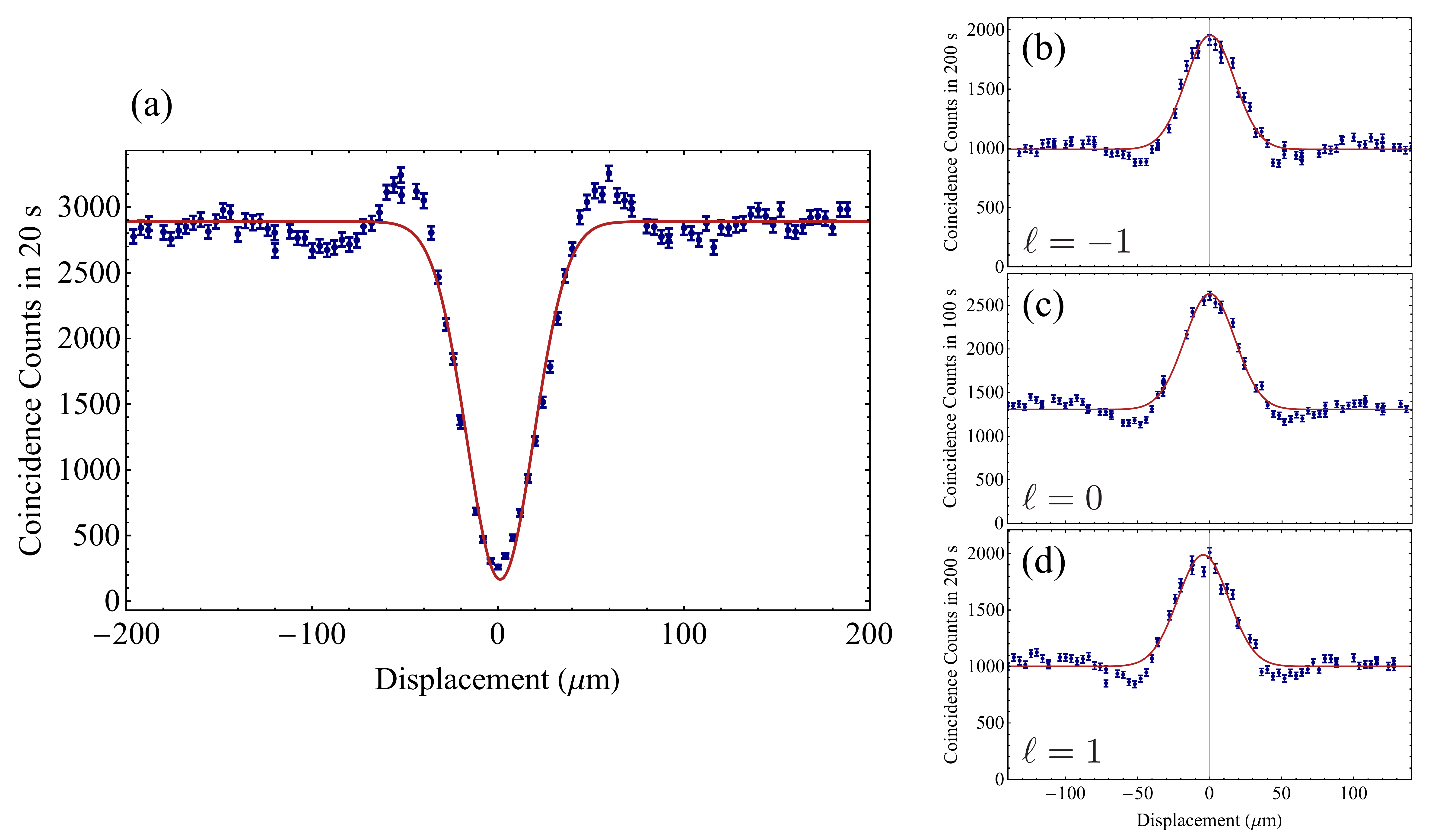}
	\caption[]{(a) Hong-Ou-Mandel interference curve for two input photons with $\ell=0$. The coincidences between the two output ports of the beam splitter is measured for various delays of one of the input photons. The experimental data points are obtained by integrating over 20 seconds and are shown along a fitted curve. A visibility of ${\cal V}=0.89 \pm 0.01$ is obtained from the fitted curve which is in good agreement with the theoretical prediction of ${\cal V}_\mathrm{th}=1$. A perfect interference of the two photons would yield a visibility of 100\%. (b)-(d) Hong-Ou-Mandel coalescence curves for input photons of $\ell=-1,0,1$ respectively (top to bottom). The indistinguishable photons will \qo{bunch} at a first beam splitter and are further split at a second beam splitter. The curve is obtained by recording the coincidences between the output ports of the second beam splitter for various delays of one of the input photons. Enhancement peaks of $R_{\ell=-1}=1.97 \pm 0.08$, $R_{\ell=0}=2.02 \pm 0.08$ and $R_{\ell=1}=1.99 \pm 0.09$ are obtained experimentally, which is in good agreement with the theoretical value of $R_\mathrm{th}=2$.}
	\label{figsm1}
	\end{center}
\end{figure}

The experimental setup can be divided in three parts: a single photon source, a Hong-Ou-Mandel (HOM) interferometer and a cloning characterization apparatus. Single photon pairs are generated by the process of spontaneous parametric downconversion at a nonlinear type-I $\beta$-barium borate (BBO) crystal illuminated by a quasi-continuous wave UV laser operating at a wavelength of $355~\mathrm{nm}$. The single photons are spatially filtered to the fundamental gaussian mode by coupling the generated pairs to single mode optical fibres, with a measured coincidence rate of $30~\mathrm{kHz}$, within a coincidence time window of $5~\mathrm{ns}$. The partner photons are each made to illuminate a spatial light modulator (SLM), in order to generate the desired photonic states, and subsequently sent at a 50:50 non-polarizing beam splitter, one at each input ports. The symmetrization method relies on the well-known two-photon interference effect at a 50:50 beam splitter first proposed by Hong, Ou and Mandel. When two indistinguishable single-photons enter a beam splitter, one into each input port, the photons will \qo{bunch} due to their bosonic nature and exit the beam splitter together through the same output port. Thus, no coincidence detections of photons exiting at different output ports will be ideally recorded. This principle is the very essence of the symmetrization method for optimal quantum cloning. The path taken by the photons generated at the nonlinear crystal to get to the beam splitter, must be equidistant for both photons of a given pair in order to observe the two-photon interference effect. This can be achieved with a precision of tens of microns using a programable translational stage, see Fig.~\ref{figsm1}~(a). 

Moreover, polarizers and interference filters are inserted in the path of each photons. The photons were then made indistinguishable in arrival time, polarization and frequency. On the other hand, the spatial modes of the photons are kept as a degree of freedom representing photonic quantum states for the UQCM. Following the HOM interference beam splitter, the bunched photons are sent to a second beam splitter separating them to further use coincidence detection. Finally, the separated output cloned photons are detected and characterized by the use of SLMs followed by single mode optical fibres, see Fig.~\ref{figsm1}~(b)-(d).

\section*{\underline{\large{Part 2}}\\ Cloning fidelities for various mutually unbiased bases:}
In quantum information, mutually unbiased bases (MUBs) are orthogonal bases such that two elements, $|\psi_i\rangle$ and $|\phi_j\rangle$, belonging to different MUBs obeys the following inner product relation,

\begin{equation}
\left| \langle \psi_i | \phi_j \rangle \right|^2 = \frac{1}{d}.
\end{equation}

An important result in quantum information is the existence of a number of MUBs for a specific dimension. In the particular case of dimensions that are powers of prime numbers, the number of MUBs is given by $d+1$. Let us now give explicit expression for each elements of every MUBs, $|\psi_i^{(\alpha)}$, where $i \in [1,d ]$ and $\alpha \in [1,d+1 ]$. The first MUB is given by the logical basis, i.e. $|\psi_i^{(1)}\rangle=|i\rangle$. Further, the elements of each MUBs with $\alpha \geq 2$ are given by the following expressions,

\begin{equation}
|\psi_n^{(\alpha)} \rangle =\frac{1}{\sqrt{d}} \sum_{j=1}^d \exp \left[ \frac{2 \pi i}{d} \left( (n-1)(d-j+1) - (\alpha-2)\left( \sum_{m=j-2}^{d+1}m \right) \right) \right] |j\rangle.
\end{equation}

\begin{figure*}[!htbp]
	\begin{center}
	\includegraphics[width=0.95\columnwidth]{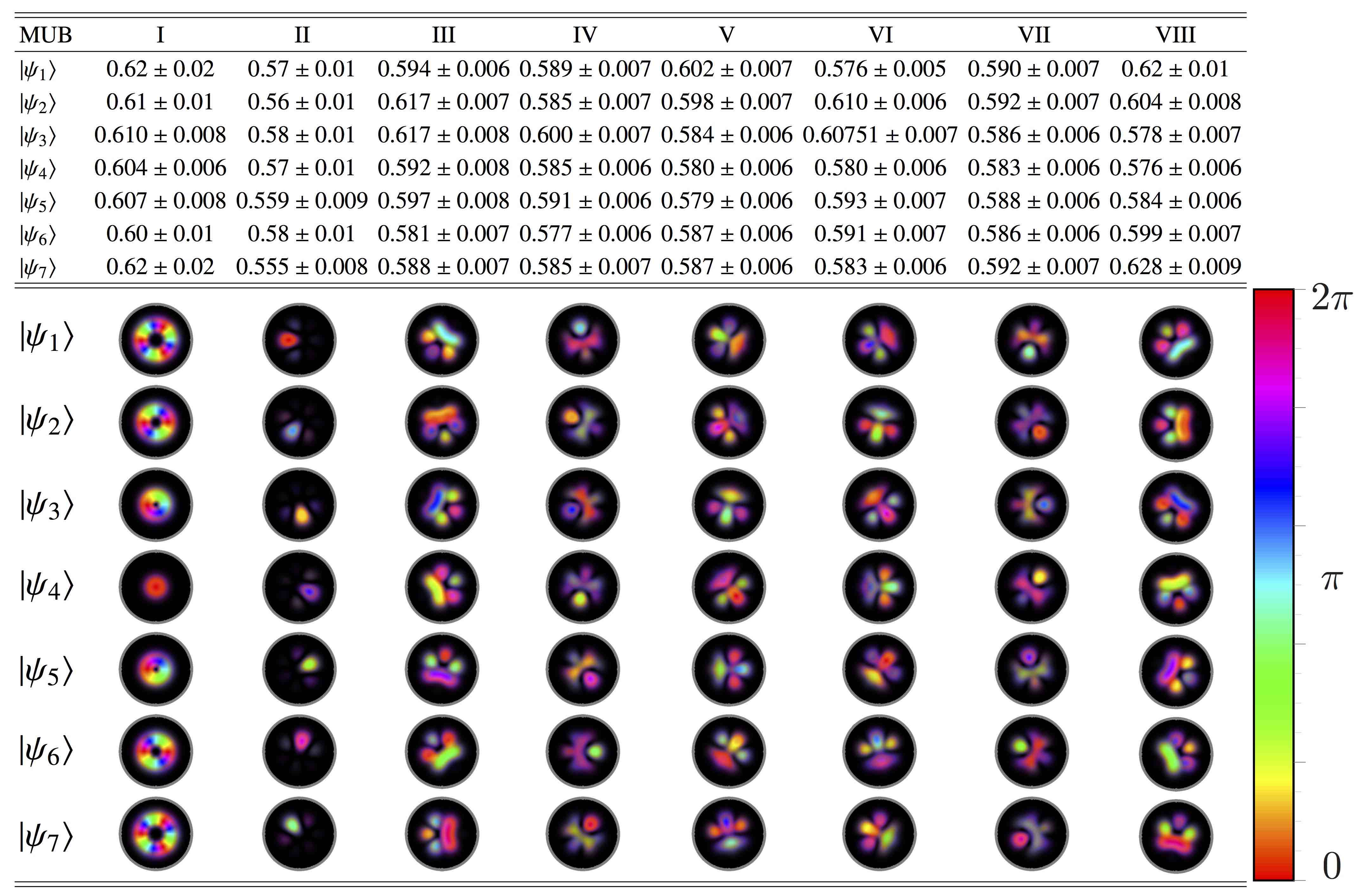}
	\caption[]{Experimental cloning fidelities for every elements of each mutually unbiased bases in dimension 7. The elements $|\psi_i \rangle$ corresponds to the input state that is to be cloned. Experimental data are shown for all $d+1$ MUBs (I)-(VIII). In the bottom, the transverse profile of each elements are shown for every MUBs. The transverse profile is illustrated by plotting the transverse phase modulated by the intensity profile, allowing one to visualize both the phase and intensity pattern at the same time.}
	\label{figsm2}
	\end{center}
\end{figure*}

In order to test the universality of our cloner, we perform cloning fidelity measurements for each elements of every MUBs, see Fig.~\ref{figsm2}.

\end{document}